\documentclass{PoS}

\newcommand{\beq}{\begin{eqnarray}}
\newcommand{\eeq}{\end{eqnarray}}

\newcommand{\pardis}{\langle {\cal M} \rangle}

\newcommand{\ie}{{\it i.e.\ }}

\newcommand{\real}{{\sf I}\kern-.12em{\sf R}}
\newcommand{\comp}{{\sf I}\kern-.50em{\sf C}}
\newcommand{\unity}{{\sf I}\kern-.54em{\sf 1}}
\def\spose#1{\hbox to 0pt{#1\hss}}
\def\ltapprox{\mathrel{\spose{\lower 3pt\hbox{$\mathchar"218$}}
 \raise 2.0pt\hbox{$\mathchar"13C$}}}

\title{Confining properties of 2--color QCD at finite density}

\ShortTitle{Confining properties of 2--color QCD at finite density}

\author{\speaker{Massimo D'Elia}, Simone Conradi and Alessio D'Alessandro\\
        \\
        E-mail: \email{delia@ge.infn.it, Simone.Conradi@ge.infn.it, Alessio.DAlessadro@ge.infn.it}}

\abstract{We study the confining properties of QCD with two colors
  across
the finite density phase transition. A disorder parameter detecting
  dual
superconductivity of the QCD vacuum is used as a probe for the 
confinement/deconfinement phase transition.}

\FullConference{XXIV International Symposium on Lattice Field Theory\\
		 July 23-28 2006\\
		 Tucson Arizona, US}

\begin{document}

\section{Introduction}

Confinement of color is a well established property of strongly
interacting matter at low temperatures and densities, even if not yet 
fully understood starting from the QCD first principles. The existence
of a high temperature phase transition to a deconfined state of matter
has been predicted by numerical simulations of lattice QCD and is
presently under investigation in Heavy Ion experiments. 
In the present study we address the question regarding the fate
of confining properties as the low temperature - high density phase 
transition is crossed: that is relevant in order to 
characterize the nature of matter in compact astrophysical objects
and more in general to understand how deconfinement at high densities compares
to what happens at high temperatures.

Numerical studies of QCD at finite density are notoriously
difficult because of the sign problem, which makes usual importance
sampling simulations unfeasible: for that reason we have 
studied the theory with 2 colors, where that problem 
is absent. No sensible differences are expected for the confining
properties of the theory  when going from $N_c = 2$ to $N_c = 3$,
where $N_c$ is the number of colors: for that reason we believe that
our study could be relevant also for real QCD.

Indications about deconfinement at high density obtained so
far
have been based on the analysis of the Polyakov loop~\cite{hands}, 
which however is not
a true order parameter for confinement in presence of dynamical fermions.
Different order parameters can be constructed in the framework of
specific mechanisms for color confinement.
One successful mechanism is that based on dual superconductivity of 
the QCD vacuum~\cite{thooft75,mandelstam,parisi}, which relates
confinement to the spontaneous breaking of a dual magnetic symmetry induced
by the condensation of magnetic monopoles: in that context 
a disorder parameter can be
developed~\cite{deldebbio:95,DiGiacomo:1997sm}  
which is the expectation value
of a magnetically charge operator, 
$\pardis$\footnote{We change the usual notation for the disorder
  operator,
$\langle \mu \rangle$, in order to avoid confusion with the notation
  for
the chemical potential.  
}, and which has been successfully tested as an order parameter for
color confinement both in the quenched theory~\cite{PaperI,PaperIII}  and in
presence of dynamical fermions~\cite{full1,full2}. Similar parameters
have been constructed elsewhere~\cite{moscow,bari1,bari2,marchetti}.

Our plan is to study the behaviour of $\pardis$ in the whole $T -
\mu$ plane, in order to characterize the confining properties 
of the various phases in the QCD phase diagram. 
In Section~\ref{definition} we will review the definition of the
disorder parameter and present our strategy for its numerical study;
some preliminary results concerning the theory with $8$ flavors of 
staggered quarks will be presented in Section~\ref{results}.

\section{The disorder parameter $\pardis$} 
\label{definition}

{${\cal M} (\vec x,t)$} is defined in the continuum as the
operator which creates a magnetic charge in $\vec x,t$ by shifting 
the quantum field by the monopole vector potential 
$\vec b_\perp (\vec y - \vec x)$
\begin{equation}
{\cal M} (\vec x, t) =  \exp \left[ i \int\!\! d\vec y \,\,\vec E_{\perp\,\,diag} (\vec y, t) \vec b_\perp (\vec y - \vec x) \right]  \, 
\end{equation}
Its expectation value, when discretized on the lattice, appears in the 
following form
\begin{eqnarray}
\label{defmu}
\pardis = \tilde{Z}/{Z} \, ;
\eeq
$Z$ is the usual QCD partition function
\beq
Z &=& \int \left( {\cal D}U \right) \, \det M (\mu)  \, e^{-\beta S}
\eeq
where $M$ is the fermionic matrix and $S$ is the pure gauge action, while
the partition function $\tilde{Z}$ is obtained from $Z$ by a change in the pure
gauge action $S \to \tilde S$ which adds the monopole field to the temporal
plaquettes  at timeslice $t$.

Being expressed as the ratio of two different partition functions, 
the numerical study of $\pardis$ is a highly non trivial task: while 
numerical methods have been recently developed which 
permit a direct determination of $\pardis$~\cite{muu1}, we will not use them in
the present study since they involve the combination of several
different Monte Carlo simulations, which in presence of dynamical
fermions could be unpractical. We will instead study, as usual, 
susceptibilities
of the disorder parameter, from which the behaviour of $\pardis$ at the 
phase transition can be inferred.

For instance, being interested in $\pardis$ as a function of $\beta$, 
as for the 
$\mu = 0$ phase transition, one usually 
measures~\cite{deldebbio:95,DiGiacomo:1997sm,PaperI}
\begin{eqnarray}
\rho = \frac{d}{d \beta} \ln \pardis = \frac{d}{d \beta} \ln \tilde Z -
\frac{d}{d \beta} \ln Z = \langle S \rangle_S -  \langle \tilde{S} \rangle_{\tilde{S}} \; 
\label{rhoferm}
\end{eqnarray}
where the subscript means the pure gauge action used for Monte Carlo sampling. 
The disorder parameter can be reconstructed from the susceptibility
$\rho$, exploiting the fact that $\pardis = 1$ at $\beta = 0$  
\begin{eqnarray} 
\label{mufromrho}
\pardis (\beta) = \exp\left(\int_0^{\beta} \rho(\beta^{\prime}) {\rm d}
\beta^{\prime}\right) \; .
\end{eqnarray}
In particular $\rho \simeq 0$ in the confined phase means $\pardis
\neq 0$,
a  sharp negative peak of $\rho$ at the phase transition implies 
a sudden drop of $\pardis$ and $\rho$ diverging
in the thermodynamical limit in the deconfined phase means that 
$\pardis$ is exactly zero beyond the phase transition.

At finite temperature and density we are interested in studying the 
behaviour of $\pardis$ in the two parameter
space $(\beta,\mu)$, where $\mu$ is the chemical potential. For that
reason we introduce the new susceptibility
\begin{eqnarray}
\rho_D \equiv \frac{d}{d \mu} \ln \pardis = 
\frac{d}{d \mu} \ln \tilde Z - \frac{d}{d \mu} \ln Z  = 
\langle N_f \rangle_{\tilde S} -  \langle N_f \rangle_{{S}} \; 
\label{rhod}
\end{eqnarray}
where $N_f$ is the quark number operator. 
The dependence of $\pardis$ on the chemical potential $\mu$ can then be 
reconstructed as follows:
\beq
\pardis (\beta,\mu) = \pardis (\beta, 0) \exp\left(\int_0^{\mu}
\rho_D(\mu^{\prime}) {\rm d} \mu^{\prime}\right) \; \nonumber
\eeq
so that, if the starting point at $\mu = 0$ is in the confined phase
($\pardis (\beta, 0) \neq 0$), 
the behaviour expected for $\rho_D (\mu)$ in correspondence of a
possible finite density deconfinement transition will be the same showed
by $\rho$ across the finite temperature transition.

\section{Numerical results}
\label{results}

We have used staggered fermions corresponding to $N_f = 8$ degenerate
continuum flavors, with bare mass $a m = 0.07$. 
Different lattices have been considered, with a fixed temporal extent $L_t = 6$
and a variable spatial size (only $L_s = 8, 16$ so far) in order
to make a finite size scaling analysis of the 
phase transition.
The critical value of $\beta$ at $\mu = 0$ is
$\beta_c \simeq 1.59$~\cite{topoldens}:  we have 
made simulations at different values of the chemical potential
$\mu$ and at a fixed value of $\beta = 1.5 <
\beta_c$,
\ie starting from the confined phase at $\mu = 0$.
An exact HMC algorithm has been used and 
standard actions both in the gluonic and in the 
fermionic sector: we have used trajectories of unit length 
and a step size variable from $10^{-2}$ to $4 \cdot 10^{-3}$:
the reduced time step was necessary in order
to keep a reasonable acceptance rate around and above 
the phase transition, 
as a consequence of the presence of small eigenvalues
of the fermion matrix. The same problem  was at the origin of a severe slowing
down around the critical value of $\mu$, which made   
the availability of the recently
installed apeNEXT facility in Rome essential
in order to carry out simulations on the larger lattice ($L_s = 16$).
Simulations on the smaller lattice $L_s = 8$ have been carried out 
both on the APEmille facility in Pisa and on the PC farm of INFN in Genova,
using there a code adapted from the publicly available version  
by the MILC collaboration.

Our results for $\rho_D$ as a function of $\mu$ are shown in Fig.~\ref{fig1}.
$\rho_D$ shows a clear peak at a critical value of the chemical
potential $\mu_c \simeq 0.3$. The peak deepens as the lattice volume 
is increased, suggesting
the presence of a true phase transition at which $\pardis$ drops
to zero and confinement (dual superconductivity) disappears.
The position of the peak is coincident with that of other
susceptibilities and we show in Fig.~\ref{fig2} the case of the Polyakov
loop: our data are still noisy and do not show
any clear size dependence, which however is not expected since the Polyakov
loop is not an order parameter for the phase transition.

\begin{figure}
\begin{center}
\vspace{0.8cm}
\includegraphics[width=0.8\textwidth]{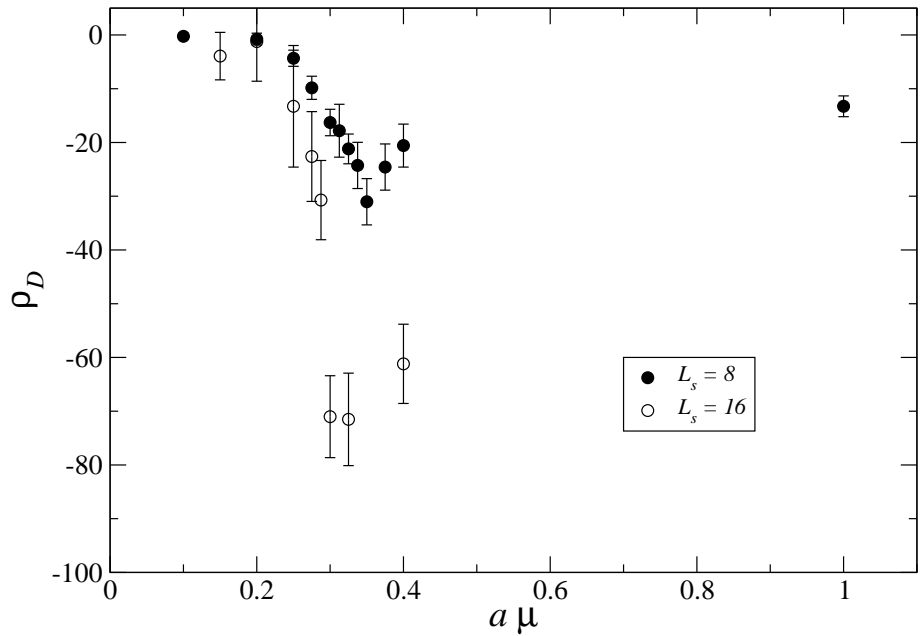}
\vspace{0.3cm}
\caption[]{Results obtained for $\rho_D$
}
\label{fig1}
\end{center}
\end{figure}

\begin{figure}
\begin{center}
\vspace{0.6cm}
\includegraphics[width=0.7\textwidth]{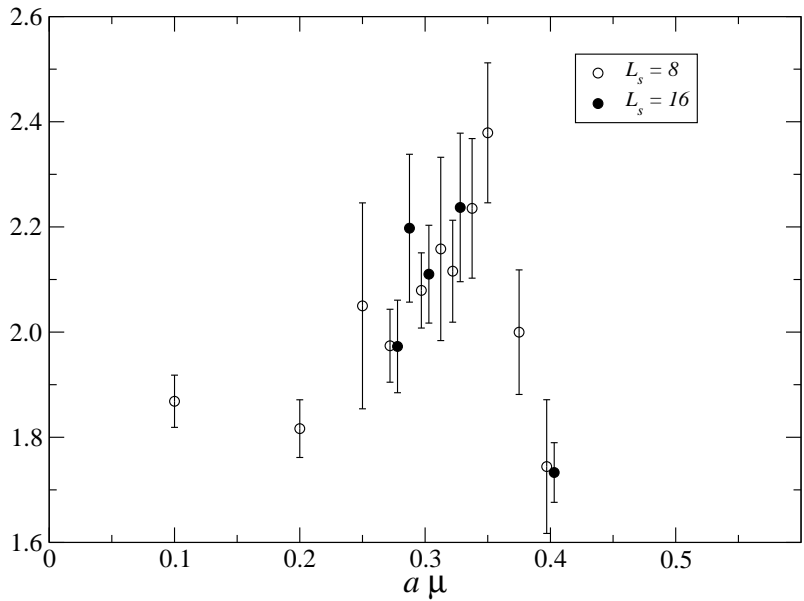}
\caption[]{Results obtained for the suscepibility of the Polyakov loop.
}
\label{fig2}
\end{center}
\end{figure}

A more detailed analysis of the disorder parameter around the phase
transition
can be carried out as follows. At fixed $T$ we can assume the
following finite size scaling behaviour for 
$\pardis$ as a function of $\mu$.
\begin{eqnarray}
\pardis = L_s^{-\beta/\nu} \Phi ( (\mu - \mu_c) L_s^{1/\nu}) \nonumber
\end{eqnarray}
from which it is easy to derive 
\begin{eqnarray} 
\rho_D = L_s^{1/\nu} \phi((\mu - \mu_c) L_s^{1/\nu}) \; 
\end{eqnarray}

Our data show a nice scaling with {$\nu \sim 0.66$}, as shown
in Fig.~\ref{fig3},
which is compatible with 
a second order phase transition in the universality class
of the $3d$ Ising model.

\section{Conclusions}
\label{conclusions}

We have made use of the disorder parameter $\pardis$ detecting dual 
superconductivity of the vacuum
to inspect the confining properties of QCD with 2 colors at finite
temperature  and density. The study has been carried our with 
$N_f = 8$ flavors of staggered fermions.
In order to analyse $\pardis$ as a function of the chemical potential 
$\mu$ we have introduced a new susceptibility $\rho_D = d/d\mu \, \ln
\pardis$, showing the presence of a phase transition at finite density
where confinement (dual superconductivity) disappears; the transition
is coincident with those signalled by other quantities such as
the Polyakov loop. A preliminary finite size scaling analysis of the
disorder parameter is compatible with a phase transition in the
universality
class of the $3d$ Ising model.

After these preliminary results, we plan in the future to make a more
extensive study of the disorder
parameter in order to characterize the whole phase diagram of the 
theory with two colors, with a particular interest in the region 
of low temperatures and high densities. To that aim,
a combined study of both susceptibilities, $\rho$ and $\rho_D$, could
be particularly useful:
indeed the knowledge of $\vec \nabla \ln \pardis$, with 
$\vec \nabla  = (d/d\beta, \, d/d\mu)$, in the whole 
$\beta - \mu$ plane, could give information
not only on the location but also on the direction of the critical
line, thus permitting a more careful study of the phase diagram.

\begin{figure}
\begin{center}
\includegraphics[width=0.73\textwidth]{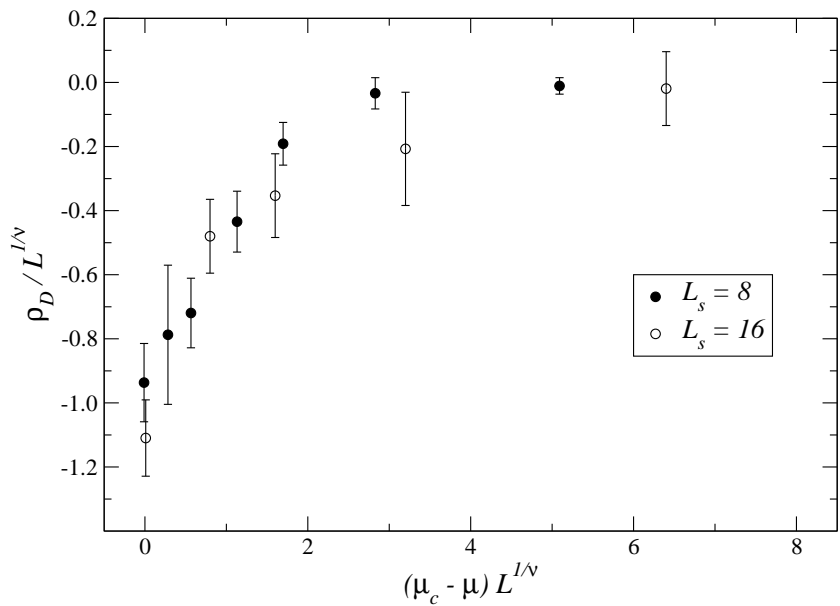}
\caption[]{Finite size scaling analysis for $\rho_D$ 
}
\label{fig3}
\end{center}
\end{figure}

\end{document}